\documentclass{aastex}
\usepackage{emulateapj5}

\begin{document}

\title{The Cosmic Background Imager}
\author{S. Padin, M.C. Shepherd, J.K. Cartwright, R.G. Keeney, B.S. Mason, T.J.
Pearson, \\A.C.S. Readhead, W.L. Schaal, J. Sievers, P.S. Udomprasert and
J.K. Yamasaki}
\affil{California Institute of Technology, MS 105-24, Pasadena, CA 91125
\\spadin@caltech.edu, mcs@astro.caltech.edu, jkc@astro.caltech.edu,
rgk@ovro.caltech.edu, bsm@astro.caltech.edu, tjp@astro.caltech.edu,
acr@astro.caltech.edu, wschaal@aol.com, js@astro.caltech.edu,
psu@astro.caltech.edu, jky@astro.caltech.edu}
\author{W.L. Holzapfel}
\affil{University of California, Department of Physics, 426 LeConte Hall, Berkeley,
CA 94720 \\swlh@cfpa.berkeley.edu}
\author{J.E. Carlstrom}
\affil{University of Chicago, Department of Astronomy and Astrophysics, 5640 S.
Ellis Ave., Chicago, IL 60637; jc@hyde.uchicago.edu}
\author{M. Joy}
\affil{NASA Marshall Space Flight Center, Dept. Space Science SD50, Huntsville, AL
35812 \\marshall.joy@msfc.nasa.gov}
\author{S.T. Myers}
\affil{National Radio Astronomy Observatory, PO Box 0, Socorro NM 87801
\\smyers@aoc.nrao.edu}
\and
\author{A. Otarola}
\affil{European Southern Observatory, Balmaceda 2536, Antofagasta, Chile
\\aotarola@eso.org}

\begin{abstract}
Design and performance details are given for the Cosmic Background Imager
(CBI), an interferometer array that is measuring the power spectrum of
fluctuations in the cosmic microwave background radiation (CMBR) for
multipoles in the range $400<l<3500$. The CBI is located at an altitude of
5000 m in the Atacama Desert in northern Chile. It is a planar synthesis
array with 13 0.9-m diameter antennas on a 6-m diameter tracking
platform. Each antenna has a cooled, low-noise receiver operating in the
26--36 GHz band. Signals are cross-correlated in an analog filterbank
correlator with ten 1 GHz bands. This allows spectral index measurements
which can be used to distinguish CMBR signals from diffuse galactic
foregrounds. A 1.2 kHz 180$^{\circ }$ phase switching scheme is used to
reject cross-talk and low-frequency pick-up in the signal processing system.
The CBI has a 3-axis mount which allows the tracking platform to be rotated
about the optical axis, providing improved $(u,v)$ coverage and a powerful
discriminant against false signals generated in the receiving electronics.
Rotating the tracking platform also permits polarization measurements when
some of the antennas are configured for the orthogonal polarization.
\end{abstract}


\section{Introduction}

The cosmic microwave background radiation (CMBR) offers us a unique view of
the early universe. Small-scale anisotropies in the CMBR contain information
about acoustic oscillations in the primordial plasma and provide a way of
measuring the fundamental cosmological parameters (Hu and White 1997; Hu et al. 1997). 
Anisotropies have been observed on angular scales from a few
degrees to $\sim \frac{1}{4}^{\circ }$ at the level of a few tens of $\mu $%
K, but measurements on smaller angular scales are important because they
can be used to further
constrain $\Omega _{b}$, $\Omega _{\Lambda }$, $h$ and the slope of the
primordial fluctuation spectrum. In this paper, we describe the Cosmic
Background Imager (CBI), an interferometer array which is measuring CMBR
anisotropies on angular scales in the range $\sim 5$'--$0.5^{\circ }$.

The goal of CMBR anisotropy measurements is to measure the angular power
spectrum of the sky intensity distribution, $P_{sky}$. On small angular
scales, this is the Fourier Transform ($FT$) of the sky temperature
autocorrelation function, $C_{sky}$. A radio telescope measures the
convolution of the sky brightness, $I_{sky}$, with the telescope beam
pattern, $B$, and the observed temperature autocorrelation function is

\noindent $C_{obs}=I_{obs}\otimes I_{obs}=(I_{sky}*B)\otimes
(I_{sky}*B)\\=(I_{sky}\otimes I_{sky})*(B\otimes
B)=C_{sky}*C_{beam}=FT(P_{sky}.W)$

\noindent where $C_{beam}$ is the autocorrelation function of the telescope beam and $%
W=FT(C_{beam})$ is called the window function for the telescope. $*$ and $%
\otimes $ indicate convolution and correlation, respectively. For an
aperture distribution $E$, the beam pattern is

\noindent $B=FT(E).FT(E)=FT(E*E)$

\noindent and if the beam pattern is symmetric

\noindent $W=FT(B\otimes B)=FT(B*B)\\=FT(FT(E*E)*FT(E*E))=(E*E)^{2}$

\noindent CMBR temperature fluctuations can be expanded in spherical harmonics
(Peebles 1993; Peacock 1999)

\noindent $\frac{\Delta T(\theta ,\phi )}{T_{o}}=\sum\limits_{l=0}^{\infty
}\sum\limits_{m=-1}^{l}a_{lm}Y_{lm}(\theta ,\phi )$

\noindent and assuming rotational symmetry, the expected value of the two-point
correlation for fields separated by angle $\theta $ on the sky is

\noindent $C_{sky}(\theta )=\frac{1}{4\pi }\sum\limits_{l=0}^{\infty
}(2l+1)C_{l}P_{l}(\cos \theta )$

\noindent where $P_{l}$ are Legendre polynomials and $C_{l}=\left\langle \left|
a_{lm}\right| ^{2}\right\rangle $. The observed correlation is

\noindent $C_{obs}(\theta )=\frac{1}{4\pi }\sum\limits_{l=0}^{\infty
}(2l+1)C_{l}P_{l}(\cos \theta )*C_{beam}\\=\frac{1}{4\pi }\sum\limits_{l=0}^{%
\infty }(2l+1)C_{l}W_{l}P_{l}(\cos \theta )$

\noindent where $W_{l}$ is the azimuthal average of $W$. Theoretical predictions and
observations are usually quoted as $\Delta T_{l}=\sqrt{l(l+1)C_{l}/2\pi }$
which is a measure of the power per log $l$. Fig. 1 shows some measurements
of CMBR anisotropy, including the first results from the CBI, along with a $%
\Lambda $-CDM model.

\vspace{3mm}
\epsfxsize=252pt
\hspace{-13pt}
\epsffile{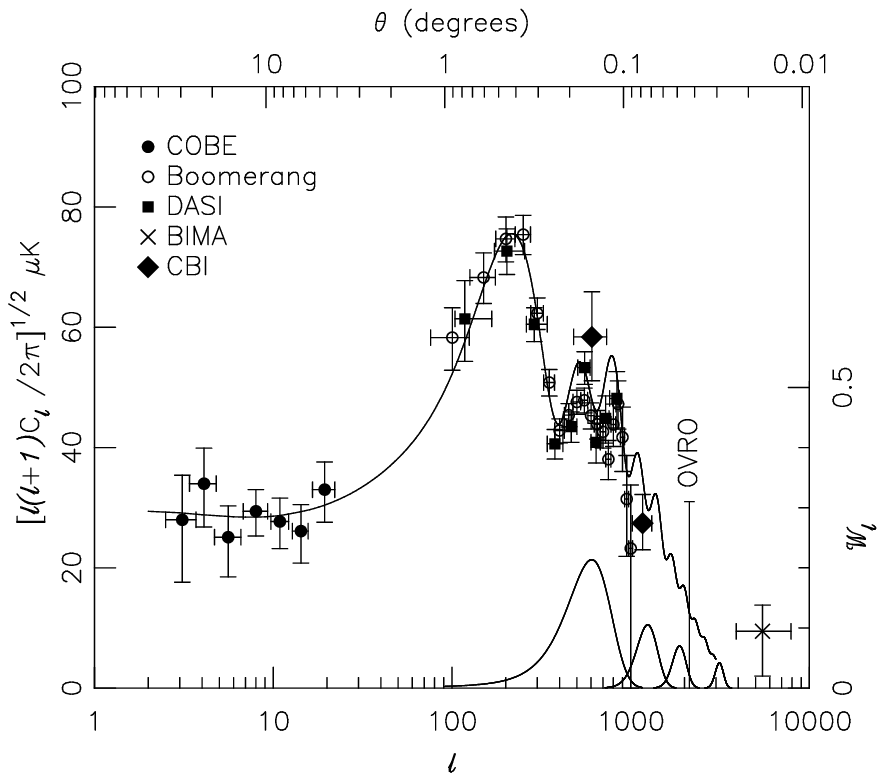}

\footnotesize
Fig. 1.--- Some CMBR anisotropy measurements, shown with a $\Lambda $%
-CDM model with $\Omega _{0}=1$, $h=0.70$, $\Omega _{b}h^{2}=0.02$ and $%
\Omega _{\Lambda }=0.66$. At bottom are window functions for 1 m, 2 m, 3 m
and 5 m baselines in the CBI. The 1 m baseline is the shortest possible with
our 0.9-m diameter antennas and the tracking platform supports baselines up
to 5.5 m. The window functions were obtained by computing the azimuthal
average of $(E*E)^{2}$ using a gaussian approximation to the aperture
distribution for a single antenna, $E_{ant}(r)=\exp -(2.3r/d)^{2}$. Data
points are from COBE (Hinshaw et al. 1996), Boomerang (Netterfield et al.
2001), DASI (Halverson et al. 2001), BIMA (Dawson et al. 2000) and CBI
(Padin et al. 2001a). The upper limit (95\% confidence) is from OVRO
(Readhead et al. 1989). The model was computed using CMBFAST (Seljak and
Zaldarriaga 1996, 1997; Zaldarriaga et al. 1998).
\normalsize
\vspace{3mm}

For an interferometer with antennas of diameter $d$ and baseline $D$, the
window function, $W_{l}$, peaks at $l\sim 2\pi D/\lambda $ and falls to zero
at $l=2\pi (D-d)/\lambda $ and $l=2\pi (D+d)/\lambda $. Fig 1 shows $W_{l}$
for several CBI baselines. $W_{l}=1$ corresponds to the thermal sensitivity
of the interferometer ($\sim $6 $\mu $K in 6 hrs for a 1 GHz band in the
CBI). The longer baseline interferometers trade sensitivity for better
resolution. They instantaneously sample a smaller azimuth range, giving
lower peaks in the azimuthally averaged window function, $W_{l}$, and lower
sensitivity for measurements of the power spectrum.

We chose to build an interferometer array for measuring the power spectrum
of CMBR fluctuations for several reasons. An interferometer provides a direct
measurement of the power in a particular range of multipoles, while in a single
dish observation the power spectrum is extracted from measurements of the sky
brightness distribution. The approaches are equivalent, but they differ in
detail and this provides an important check. Single dish observations which
use a switching scheme can reject the total power, and hence 1/f noise from
the receiver, only at frequencies below the switching frequency, but an
interferometer has the advantage of no total power response on all timescales.
An interferometer is a good way of rejecting atmospheric noise, because
atmospheric brightness fluctuations tend to wash out as they drift through
the fringe pattern (Webster 1994). An interferometer also permits measurements
on small angular scales without large optics. The imaging capability of an
interferometer is very important for testing, and we used this
extensively to help us understand the ground contribution in our observations.

The CBI operates in the 26--36 GHz band, which was chosen as a compromise
between receiver sensitivity and contamination from foreground sources. At
lower frequencies, point sources and galactic synchrotron and free-free
emission are a serious problem, while at higher frequencies, receiver noise
temperatures, emission from interstellar dust and atmospheric noise are all
increasing. The 325 MHz Westerbork Northern Sky Survey (Rengelink et al.
1997) shows fluctuations in synchrotron emission at $\sim $2\% of the total
intensity on scales of a few arcminutes. In the CBI fields, the 325 MHz
total intensity is $\sim $30 K, so we expect fluctuations in the synchrotron
emission to contribute only $\sim $3 $\mu $K at 31 GHz. The free-free
contribution is limited by H$\alpha $ observations (Gaustad et al. 1996) to
a few $\mu $K, but the anomalous $\beta \sim -2$ (where $\Delta T\propto \nu
^{\beta }$) emission from the NCP measured by Leitch et al. (1997) is a
potential problem. This anomalous emission is highly correlated with IRAS
100 $\mu $m emission (Kogut 1999), so we have chosen CBI fields in regions
with 100 $\mu $m brightness $<$1 MJy sr$^{-1}$. In these regions, emission
from dust is just a few $\mu $K and fluctuations on arcminute scales are
substantially less. The diffuse foreground contribution can be extracted
using spectral index measurements (Brandt et al. 1994) and for this purpose,
the CBI has ten 1-GHz bandwidth correlators covering the 26--36 GHz band. The
wide bandwidth also gives high sensitivity. In the case of strong foreground
contamination, the sensitivity penalty in extracting the foreground is
severe, but our initial CBI observations show no significant contamination.

At high $l$, point sources are a serious problem for the CBI. The power
spectrum of CMBR fluctuations falls off rapidly with $l$, while the
point source contribution increases roughly as $l^{2}$. For $l>1500$, CBI
measurements are strongly contaminated, so we measure the flux densities of
point sources in the CBI fields using a 26--34 GHz Dicke switched radiometer
on the 40 m telescope at the Owens Valley Radio Observatory (OVRO). Point
source contamination would be significantly lower in the 90 GHz band, and
improvements in 90 GHz receivers make this an attractive option for a
future upgrade to the CBI.

\vspace{6mm}
\section{The CBI site}

The CBI is located at an altitude of 5000 m in the Andes near Cerro
Chajnantor, 40 km east of San Pedro de Atacama, Chile. This choice was based
largely on our experience with CMBR observations with the OVRO 5 m telescope
at 32 GHz and $l\sim 600$ (Leitch et al. 2000). The sensitivity of the 5 m
telescope is limited by atmospheric noise for all but 2 or 3 nights per
year, and for the CBI to achieve the thermal noise limit it had to be at a
site with about four times less atmospheric noise than OVRO. This implies an
altitude of $\sim $4000 m, but the estimate is very rough because the noise
depends on the distribution of water vapor in the atmosphere (Lay and
Halverson 2000). Site testing is the only way to resolve this issue, but a
testing phase was not appropriate for a small, short-term project like the
CBI, so we had to choose an established site. This narrowed the options to
Chajnantor, Mauna Kea and the South Pole. Point source subtraction is a
serious problem for the CBI at the South Pole, because there are no
high-resolution 30 GHz telescopes available for deep point source surveys in
the Southern Hemisphere. Mauna Kea would have involved a long delay and
considerable expense, so we chose Chajnantor. Chajnantor is a good
millimeter-wave site, has fairly good access and nearby towns provide some
infrastructure. However, the environment is aggressive and the combination
of high altitude, snow, high winds and low temperatures makes it a difficult
place to work. During the period November 1999 to May 2001, we lost about
one third of the observing time as storms passed through. (We also had an
eruption of Volcan Lascar, $\sim $40 km south of the CBI, but fortunately
that turned out to be all smoke and no fire!) During the rest of the time,
conditions were excellent and the CBI operated at essentially the thermal
noise limit.

The Chajnantor site has no infrastructure, so the CBI has its own power
plant consisting of a pair of identical 300 kVA diesel generators (Atlas
Copco). These are de-rated $\sim $50\% for operation at 5000 m. While one
generator is providing power for the telescope, the other is a standby which
will start automatically if the first generator fails. We switch generators
every week to permit routine maintenance. Fuel consumption is $\sim $15 m$^{3}$
per month, and we can store up to 40 m$^{3}$ on the site. (A large fuel stock
is important during the winter, because access for a fuel truck may not be possible
for 4--6 weeks following a snow storm.) The buildings at the CBI site are
standard ISO shipping containers which are insulated and fitted with heaters
and air conditioners, power and lights. The site has a control room,
laboratory, 2 bedrooms, bathroom, machine shop and several unfitted storage
containers. During routine operations, we usually sleep at the CBI base in
San Pedro de Atacama, but stays of 3--4
days at the site are common when the road is blocked with snow.
The entire development laboratory for the CBI was shipped with
the instrument, so we have good facilities for making repairs and adding new
equipment. The oxygen concentration in the control room, laboratory and 2
bedrooms is increased to $\sim $26\% to provide an equivalent altitude of $%
\sim $3500 m (West 1995; Cudaback 1984). Oxygen for the containers is
extracted from the outside air using molecular sieves (AirSep Corp.). A 40
ft container requires $\sim $20 l min$^{-1}$ of oxygen and provides a
comfortable working environment for 2 or 3 people. For outside work, we use
portable oxygen systems comprising a 415 l oxygen bottle, a 1--4 l min$^{-1}$
demand regulator and a nasal cannula (Chad Therapeutics Inc.). These systems
are vital to the CBI operation because they enable us to solve complex
engineering problems on the telescope.

\section{Array design considerations}

The high sensitivity requirements for CMBR observations favor an array with
a high filling factor and an attractive option is an instrument which can be
equipped with antennas of different sizes. This permits observations on a
range of baselines, but always with a fairly close-packed array. In this
scheme, the antenna diameter is equal to the shortest baseline, which is in
turn set by the smallest $l$ of interest. The instantaneous resolution in $l$
corresponds to about half the antenna diameter. Mosaicing (Cornwell et al.
1993) allows the instrument to sample smaller $l$ and improves the
resolution in $l$. We use this approach to give a factor 2 or 3 improvement
in $l$-resolution. The CBI mount was designed to support different array
configurations, with baselines up to 5.5 m, and different antenna diameters,
so that we could match the sensitivity, $l$-range and resolution of the
instrument to a particular part of the power spectrum of CMBR fluctuations.
The array currently has 0.9-m diameter antennas and a minimum baseline of 1
m, giving an $l$-range of $\sim $600--3500 and an $l$-resolution of $\sim $%
400 (FWHM), without mosaicing. We have used the antennas in fairly
close-packed configurations for maximum sensitivity at low $l$, and in a
ring-like configuration for fairly uniform sensitivity over the full $l$%
-range of the instrument.

The power spectrum of CMBR fluctuations drops off rapidly for $l>300$, so
all baselines in the CBI show a decreasing CMBR signal with increasing
frequency. Galactic synchrotron and free-free emission have similar
signatures because of their power spectra. Breaking the degeneracy between
these effects requires good sampling in $l$ and is one of the key
constraints in choosing a CBI configuration. An alternative approach is to
choose baselines that sample the same, or very similar, values of $l$ at a
different frequencies, and we have done this in some CBI configurations.

In a close-packed array, cross-talk between the antennas can be a serious
problem. Noise emitted from the input of a receiver can scatter into an
adjacent antenna and cause a false signal at the correlator output, as shown
in Fig. 2. The false signal limits the sensitivity of the instrument because
it is similar to the signals we are trying to measure. In Fig. 2, receiver $x
$ has noise temperature $T_{r}$ and emits, from its input, noise power $%
pT_{r}$ which is correlated with the receiver noise. The coupling between
the antennas is $c$ and power $pcT_{r}$ is coupled into antenna $y$, causing
a false signal with maximum amplitude $T_{r}\sqrt{pc}$ at the correlator
output. Since the coupled noise arrives at the correlator with a delay
error, the false signal is reduced by a factor sinc$(\pi \tau \Delta \nu )$,
where $\Delta \nu $ is the bandwidth of the signals at the correlator inputs
and $\tau $ is the delay error (Thompson and D'Addario 1982). The false
signal is then $s=T_{r}\sqrt{pc}\ $sinc$(\pi \tau \Delta \nu )$. This
expression is for noise coupling in just one direction, but in practice the
process occurs in both directions. If the receivers are identical, the
signals for the two directions are complex conjugates, so the correlator
output is always real but the amplitude can be anywhere in the range 0 to $2s
$, depending on the coupling path length. In the CBI, $T_{r}\sim 20$ K, $%
\tau \sim 6$ ns and $\Delta \nu =1$ GHz so the false signal could be as big
as $2\sqrt{c}$ K. For false signals $<$1 $\mu $K, the inter-antenna coupling
should be $<-126$ dB. This is a very stringent requirement and is the key
parameter driving the design of the antennas in the CBI.

\vspace{3mm}
\epsfxsize=126pt
\hspace{50pt}
\epsffile{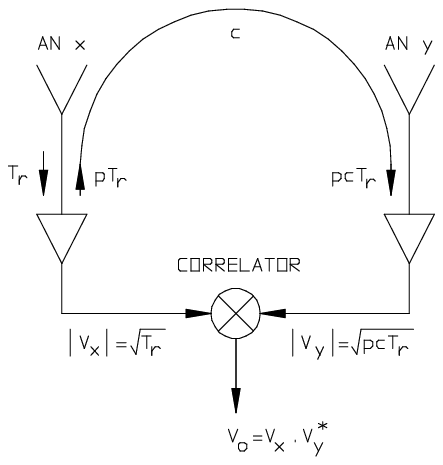}

\footnotesize
Fig. 2.---Cross-talk between antennas in an array. $T_{r}$ is the
receiver noise, $p$ is the correlation coefficient between the outgoing and
receiver noises, $pT_{r}$ is the outgoing receiver noise which is correlated
with $T_{r}$ and $c$ is the coupling between the antennas. (If we transmit $%
P_{t}$ from antenna $x$ and receive $P_{r}$ at antenna $y$, then $c=\frac{%
P_{r}}{P_{t}}$). The correlator is a complex multiplier (Thompson et al.
1986). 
\normalsize
\vspace{3mm}

The cross-talk problem also determined the type of synthesis array we chose
for the CBI. In a conventional tracking array, the antennas move with
respect to each other, so the cross-talk varies continuously during an
observation. This does tend to wash out the cross-talk, but any residual is
hard to measure. To avoid this problem, we chose a planar array for the CBI.
The antennas are all mounted on a rigid tracking platform so the cross-talk
is constant. The platform can be rotated about the optical axis, which
allows us to measure cross-talk (and other constant false signals) generated
anywhere in the instrument; the false signals rotate with the array, while
signals from the sky do not. The additional rotation axis increases the
mechanical complexity of the telescope mount, but the signal processing in a
planar array is significantly simpler than for a tracking array because
there are no fringe rotators or tracking delays. The rotating platform can
be used to improve the $(u,v)$ coverage for an observation and allows us to
track a field in parallactic angle, which keeps the response to linearly
polarized foregrounds constant. (The receivers in the CBI are configured for
circular polarization, but they have $\sim $10\% response to linear
polarization at the band edges.) Rotating about the optical axis also
permits polarization measurements when some of the antennas in the array are
configured for the orthogonal polarization. The instrumental contribution to
the response of a cross polarized interferometer rotates with the array, but
the contribution from a polarized source is constant if the array tracks the
parallactic angle (Conway and Kronberg 1969). Rotating the array relative to
the parallactic angle allows the source and instrumental contributions to be
separated. The combination of a rotating antenna platform and
a configuration with redundant baselines is a powerful tool for identifying
false signals, and this feature was used extensively during testing of the
CBI.

\section{The telescope mount}

\vspace{3mm}
\epsfxsize=252pt
\hspace{-13pt}
\epsffile{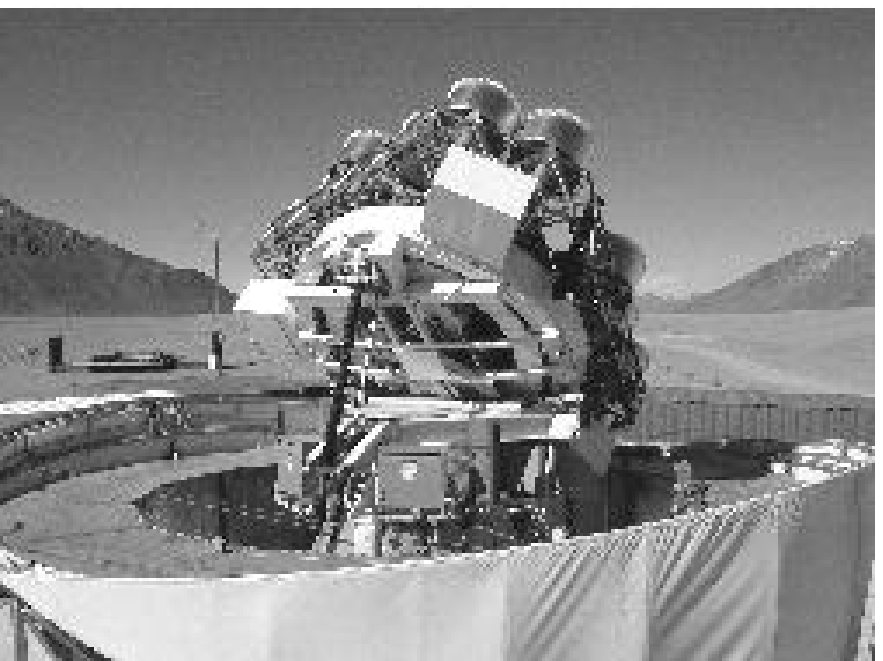}

\footnotesize
Fig. 3.---Rear view of the CBI. The struts underneath the elevation
platform support the parallactic angle cable wrap and the elevation drive
motor, which is at the top of the ballscrew. The ballscrew nut is attached
to the forks at the back of the azimuth platform. The cables and hoses at
the bottom of the elevation platform are part of the elevation cable wrap,
which is just behind the elevation axis. The three white boxes attached to
the back of the antenna platform contain the signal processing and drive
control electronics and uninterruptable power supplies. The 15 cm refractor
used for pointing measurements is to the right of the uppermost electronics
box.
\normalsize
\vspace{3mm}

\epsfxsize=252pt
\hspace{-13pt}
\epsffile{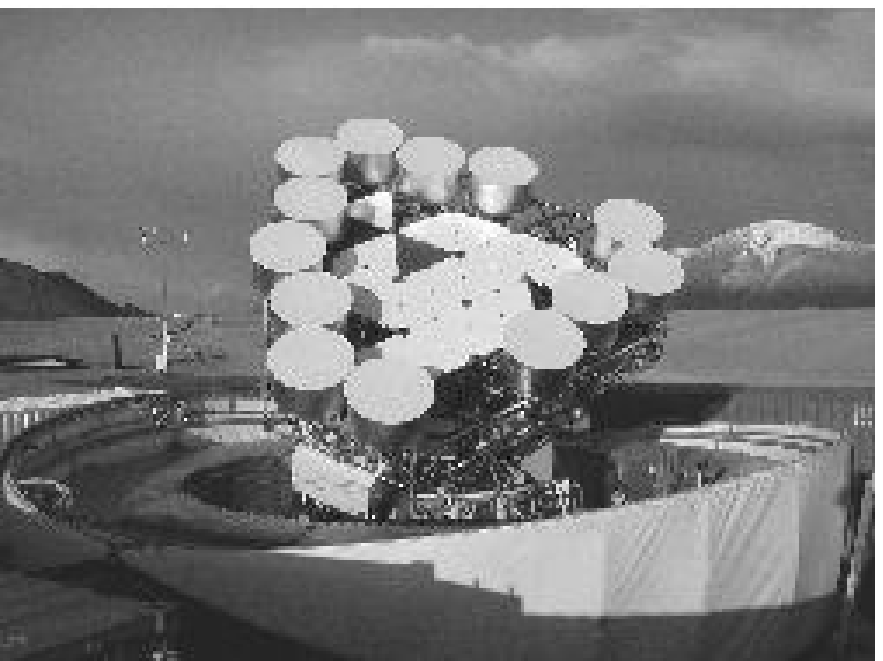}

\footnotesize
Fig. 4.---Front view of the CBI with the antennas in a ring
configuration. This permits easy access to the receivers and was used for
testing the CBI and for our first measurements of the power spectrum of CMBR
fluctuations. The helium compressors for the refrigerators are at the front
of the telescope, on a frame attached to the azimuth platform.
\normalsize
\vspace{3mm}

The design of the CBI mount was driven by the conflicting requirements of
high pointing accuracy and the need to make a small, fairly lightweight
structure that could be transported to a remote observing site. The main
effect of pointing errors is to limit the accuracy with which we can
subtract point sources from our observations. For any baseline in the
instrument, the correlator output, in temperature units, due to a point
source of flux density $S$ is $T\sim S\lambda ^{2}/2k\Omega _{p}$ where $%
\Omega _{p}$ is the primary beam solid angle. $\Omega _{p}\sim \left(
\lambda /d\right) ^{2}$, so $T\sim Sd^{2}/2k$. A small pointing error $\zeta 
$ changes the phase of the correlator output by up to $\Delta \psi =2\pi
D\zeta /\lambda $ so the maximum residual after subtracting a point source
of flux density $S$ is $\Delta S=2\pi SD\zeta /\lambda $. The residual in
temperature units is $\Delta T\sim \pi Sd^{2}D\zeta /k\lambda $. For a 5.5 m
baseline in the CBI, $\Delta T(\mu $K$)\sim 5S($Jy$)\zeta $(arcsec). The
brightest point source in a CBI field is $\sim $200 mJy, so for peak
temperature residuals of just a few $\mu $K, we require pointing errors of
just a few arcseconds. Pointing errors have the biggest effect on the longer
baselines where, unfortunately, the microwave background signal is weaker
because the power spectrum is falling off rapidly. For measurements on the
shorter CBI baselines, the pointing requirements can be relaxed by about an
order of magnitude. Mosaicing also requires good pointing, with errors less
than a few percent of the primary beamwidth (Cornwell et al. 1993). However,
this is not a severe constraint in the CBI because the primary beamwidth is $%
\sim $45' so pointing errors $<$1' do not seriously limit the dynamic range
and fidelity of mosaics.

The conventional approach to achieving arcsecond pointing for a 6 m
telescope is a mount that is balanced about all axes. However, this
significantly increases the mass and size of the instrument, making
transportation to the observing site more difficult. The CBI mount
(fabricated by L \& F Industries) has no elevation counterweight and weighs
only $\sim $19,000 kg, which was light enough for us to transport it to Chile
without any disassembly. Both the elevation and azimuth axes are unbalanced,
but the elevation range is limited to $40^{\circ }<el<89^{\circ }$ to reduce
the effect of deformations. Our observations of the CMBR are made at fairly
high elevation to reduce ground pickup and atmospheric effects, so the
limited elevation range is not a great disadvantage. The CBI mount is shown
in Figs. 3 and 4.

\begin{figure*}
\figurenum{5}
\epsfxsize=521pt
\epsfbox{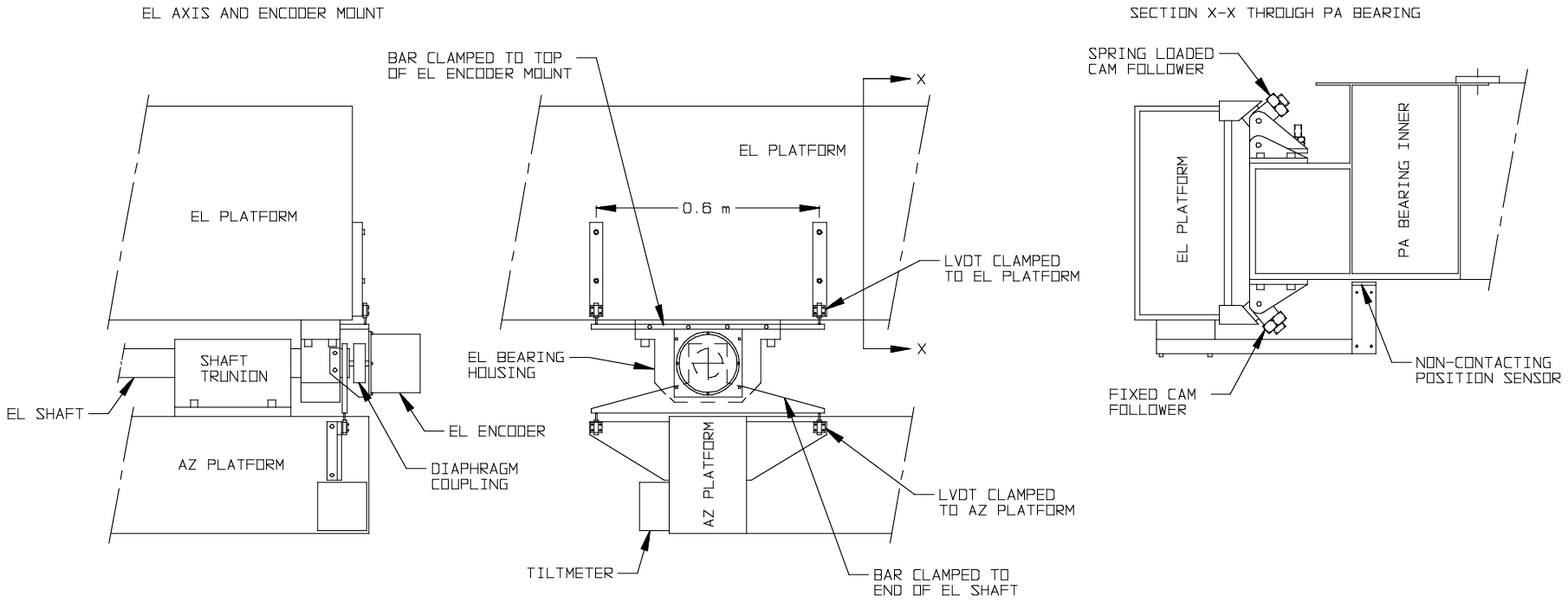}
\caption{Elevation position measurement system. The linear
displacement transducers (RDP Electrosense LVDTs) clamped to the azimuth
platform measure the rotation of a 0.6-m long bar attached to the end of the
elevation shaft. A 1'' rotation causes a 1.5 $\mu $m change in the position
of the LVDT probes. This is a very small displacement, so we make a
differential measurement using a pair of probes, which gives high immunity
to temperature variations and large-scale distortions in the telescope
structure. The probe tips contact small glass plates glued to the ends of
the bar to ensure that surface defects do not affect the measurement. The
LVDTs clamped to the elevation platform measure the rotation of a bar
attached to the encoder mount, which is in turn bolted to the elevation
bearing housing. We could significantly reduce the rotation of the end of
the elevation shaft (and probably eliminate the rotation measurement) by
adding a second trunion on the other side of the bearing housing, but this
would be difficult with the current azimuth platform design. The parallactic
angle bearing has 4 non-contacting position sensors (RDP Electrosense
capacitive sensors) equally spaced round the circumference. We measure the
difference between diametrically opposed sensors and compare with
measurements made with the telescope at the zenith. We could reduce the
elevation dependent tilt of the parallactic angle bearing by increasing the
spring force on the top cam followers. However, this increases the torque
required to turn the bearing because we use cylindrical cam followers
running on conical tracks, so the cam followers must slip as the bearing
turns.}
\end{figure*} 

The azimuth axis rotates on a 1.7-m diameter ball bearing which is located
at the bottom of the azimuth platform (see Fig. 3). The bearing has a high
preload and the change in the center of gravity of the telescope with
elevation causes a maximum axial tilt of $\sim $10''. In addition, there is $%
\sim $10'' tilt due to deformation in the base of the teepee which supports
the azimuth bearing. We measure and continuously correct the azimuth tilt
using a tiltmeter (Applied Geomechanics Inc.) mounted on the azimuth
platform in the plane of the azimuth bearing. A 22-bit absolute encoder
(BEI), mounted just below the tiltmeter, measures the position of the azimuth
axis. The encoder shaft is connected to a post which is attached to the base
of the teepee. The azimuth axis is driven by a pair of motors with $\sim $%
200 Nm torque offset. The motor pinions engage a ring gear on the inside of
the azimuth bearing and the gear ratio is only 16:1 so the axis is stiff and
fast. We use an azimuth slew rate of $\sim $1 turn min$^{-1}$ and while
faster slews are possible, this matches the speed of the other axes. The
azimuth (and elevation) drive motors are water-cooled brushless DC motors
similar to those used on the Submillimeter Array. They have a peak torque
output of $\sim $1500 Nm and the motor casing contains a disk brake and a
19-bit encoder which measures the position of the rotor for commutation.

The elevation axis rotates on a pair of 7.5-cm diameter tapered roller
bearings. These are standard commercial-grade bearings which we re-housed to
provide adjustable preload. The elevation axis is driven by a 7.5-cm
diameter ball-screw attached to the elevation platform, with a nut attached
to the back of the azimuth platform (see Fig. 3). The ball-screw is clamped
directly to the rotor in the elevation motor and since the axis is
unbalanced, there is no backlash. To minimize the length of the ball-screw,
and hence the height of the mount, we positioned the motor and nut so that
the tail of the ball screw moves rapidly away from the teepee as the
telescope approaches the zenith. As a result, the elevation gear ratio
increases from $\sim $900 at $el=40^{\circ }$ to $\sim $950 at $el=75^{\circ
}$ and then drops rapidly to $\sim $800 at the zenith. The disadvantage of
this approach is that the elevation servo parameters must be adjusted to
compensate for the changing gear ratio. Since the elevation axis is
unbalanced, a failure in the drive is potentially catastrophic, so we have
two braking systems: an electromechanical disk brake inside the motor casing
and a hydraulic brake mounted alongside the ball screw. Both brakes engage
in the event of a failure (e.g. open servo loop, power failure, drive speed
too high or telescope at a position limit). The hydraulic brake can also be
used to lower the elevation platform if the elevation drive motor fails.

In the CBI, the elevation shaft is fixed and the elevation bearing housings,
which are attached to the elevation platform, rotate. A 22-bit absolute
encoder is attached to one of the bearing housings and the encoder shaft is
connected to the end of the elevation shaft. The load on the elevation shaft
and bearing housings changes with elevation. This causes the bearing
housings to rotate $\sim $100'' relative to the elevation platform, and the
end of the elevation shaft rotates $\sim $15''. Both rotations affect the
elevation encoder reading. In addition to the rotations, there is a $\sim $%
25 $\mu $m deflection in the ends of the elevation shaft and this causes a
radial misalignment with the encoder axis. The misalignment is taken up by a
15-cm long diaphragm coupling with 3 spring plates. The coupling is folded 3
times so its insertion length is only 5 cm and this permits a very compact,
and hence stiff, elevation encoder mount. The rotation of the bearing
housings is repeatable, at the level of a few arcseconds, and can be
included in the flexure term in the pointing model, but the elevation shaft
rotation shows some hysteresis. The elevation bearing housings also rotate $%
\sim $30'' as the antenna platform is rotated. This is due to deformations
in the parallactic angle bearing and is repeatable at the level of a few
arcseconds. We measure the rotations of the bearing housing and elevation
shaft, at the elevation encoder, using linear displacement transducers as
shown in Fig. 5, and continuously apply corrections to the requested
elevation. This references the outside of the fairly rigid parallactic angle
bearing to the azimuth platform. A second tiltmeter, mounted near the
transducers which measure the rotation of the elevation shaft, can be used
to directly reference the outside of the parallactic angle bearing to the
local gravity vector. This corrects deformation in the front of the azimuth
platform, which is $\sim $10'' over the full elevation range of the CBI.
However, the correction cannot be applied continuously because the tiltmeter
is $\sim $2 m from the azimuth axis and the centripetal acceleration during
tracking gives unstable tiltmeter readings. To read this tiltmeter, we must
stop tracking for a few seconds, but since the tilt varies slowly this only
needs to be done every $\sim $10 min.

The antenna platform rotates on 40 7.5-cm diameter cam followers which run
on a pair of 3.5-m diameter ground tracks (see Fig. 5). The lower track is
welded to the elevation platform and is the reference for the parallactic
angle axis. The 20 cam followers which run on the lower track can
articulate, but their radial positions are fixed. The upper 20 cam followers
are spring loaded against the top track and they push the inner part of the
bearing against the lower track. At the zenith, the lower track deviates $%
\sim $18 $\mu $m from a circle, so the maximum axial tilt in the bearing is $%
\sim $2''. There is an additional elevation dependent tilt due to the cam
followers sliding radially on the tracks. The effect is mainly in the
elevation direction and is $\sim $3'' over the full elevation range of the
CBI. We measure this tilt using non-contacting position sensors as shown in
Fig. 5, but since the effect is small we have not yet included it as a
real-time pointing correction. Three steel drive wheels mounted on the
elevation platform turn the inner section of the parallactic angle bearing.
Each wheel is driven by a stepper motor through a 50:1 worm gearbox with a
clutch to synchronize the drives. The required pointing accuracy for the
parallactic angle axis is only $\sim $1' so we made no attempt to eliminate
backlash in the drive. The position of the parallactic angle axis is
measured by a 19-bit absolute encoder (Heidenhain Corp.).

The antenna platform is a hexagonal space frame $\sim $6 m in diameter. The
frame is 0.75 m thick with 1 m equilateral triangles on the top and bottom
surfaces. It is made from 87 identical rectangular sections which are pinned
to rods with a finned node pressed onto each end. Gravitational deformation
in the platform, over the full elevation range, with the antennas in a symmetric
configuration, is $\sim $%
50 $\mu $m which corresponds to a tilt of $\sim $2''. The entire CBI mount,
including the antenna platform, is made of steel so the whole structure has
the same expansion coefficient. This is important for operation at the
Chajnantor site because the diurnal temperature variation is $\sim $20 K and
cooling after sunset is rapid, with most of the temperature variation
occurring in 1--2 hrs. (Heating after sunrise is slower, typically $\sim $4
hrs.)

The pointing model for the CBI is determined from observations of stars
using a 15 cm refractor and CCD camera mounted near the edge of the antenna
platform. We typically determine the collimation terms for the optical
telescope by measuring the pointing errors as a function of antenna platform
rotation. Then we measure the pointing errors for $\sim $25 stars, uniformly
spread over the sky, to determine the encoder offsets, axis tilts (typically 30'')
and flexure ($\sim $30'' over the full elevation range). A model fit to the
pointing measurements typically gives $\sim $3''
rms and $\sim $6'' peak residuals, with az cos(el) and elevation residuals
of $\sim $2'' rms. Variations in the pointing model during the night, and
from night to night, are mainly in the encoder offsets, at the level of a
few arcseconds. These variations are partly due to thermal gradients and
partly due to non-repeatable changes in the telescope structure. The pointing
performance could be improved by using the 15 cm refractor for guiding, but
this would restrict observations to clear nights. In addition
to the 15 cm refractor, we have two 5 cm refractors with CCD cameras, one
mounted at the center of the antenna platform and one on the elevation
platform, just above the elevation encoder. Pointing measurements with all 3
optical telescopes allow us to measure deformations in the antenna platform
and parallactic angle bearing, which are at the level of a few arcseconds.
Short-term tracking errors are typically $%
\sim $3'' peak in azimuth and elevation. The dominant error has a period of
one motor commutation cycle and is due to gain and offset errors in the
drive amplifiers and torque ripple in the motors. Since the azimuth drive
motors have a high torque offset and the elevation platform is unbalanced,
the azimuth and elevation axes are very stiff and we do not see any
degradation in tracking with wind gusts up to $\sim $15 m s$^{-1}$.

The CBI drives are controlled by a PMAC computer (Delta Tau Data Systems
Inc.). This is designed for controlling multi-axis machine tools, but it
provides many of the functions required for telescope control. The PMAC
receives requested positions from the CBI real-time control system every
second. It moves the axes as requested, commutating the motors, checking
position, velocity and acceleration limits and monitoring the status of
amplifiers, motors and encoders. It also sets the servo loop parameters,
which are different for slewing and tracking and, in the case of the
elevation axis, vary with axis position. The speed of the axes is controlled
by a crystal oscillator and a trigger pulse from a GPS clock is used to
synchronize the PMAC each time a new source is observed.

The azimuth range of the CBI is $-180^{\circ }<az<270^{\circ }$, which was
chosen to avoid long slews between the target source and calibrators for
most observations. The parallactic angle axis also has a range of 1$\frac{1}{%
4}$ turns, but some long slews are unavoidable because in addition to
tracking in parallactic angle, we also rotate the antenna platform to
improve the $(u,v)$ coverage. The azimuth cable wrap is inside the teepee
and is of the coiled-spring design. This wrap carries power, cooling water
for helium compressors and electronics and optical fibers for communications
with the telescope control computer. The parallactic angle cable wrap is
more complicated because it has to fit inside the bearing structure while
carrying 26 helium lines for the receivers, power, cooling water and many
drive control cables. The wrap overlays 42 cables in 3 layers on a central
drum, with the cables supported on 3 satellite drums, kept in tension by
coiled springs. A key advantage of this design is that the cables in the
moving part of the wrap are only $\sim $3 m long (compared with $\sim $15 m
for the coiled-spring azimuth wrap). This helps to minimize the pressure
drop in the helium lines in the wrap.

The CBI is housed in a retractable dome (manufactured by American Space
Frames Inc.) which protects the instrument and workers from the weather. The
dome is a 12-m diameter hemispherical steel space frame covered with
polyethylene cloth (see Fig. 6). It has 8 segments which nest inside each
other, 4 on each side, when the dome is open (see Figs. 3 and 4). To close
the dome, the top 4 segments are raised on cradles and then wire ropes pull
the top 2 dome segments together and up. As a segment rises, tangs on the
bottom engage the top of the next segment in. A pneumatic clamp at the top
of the dome holds the 2 halves of the hemisphere together and the bottom
segments are clamped to the top of the walkway that surrounds the telescope.
The walkway provides easy access to the electronics boxes on the antenna
platform. The dome is a fairly light, flexible structure and the $\sim $5 cm
gaps between segments prevent large pressure differences building up, even in
very high winds. The structure has withstood storms lasting 3 days with wind
speeds up to 45 m s$^{-1}$. Under these conditions, the cloth segments tend
to pull away from the steel frame. This reduces the cross-section of the
segments and limits the wind loading, preventing serious damage to the steel
structure. Usually, we do not observe if the wind speed is much above $\sim $%
15 m s$^{-1}$ because it is difficult to open and close the dome in strong
gusty wind.

\vspace{3mm}
\epsfxsize=252pt
\hspace{-13pt}
\epsffile{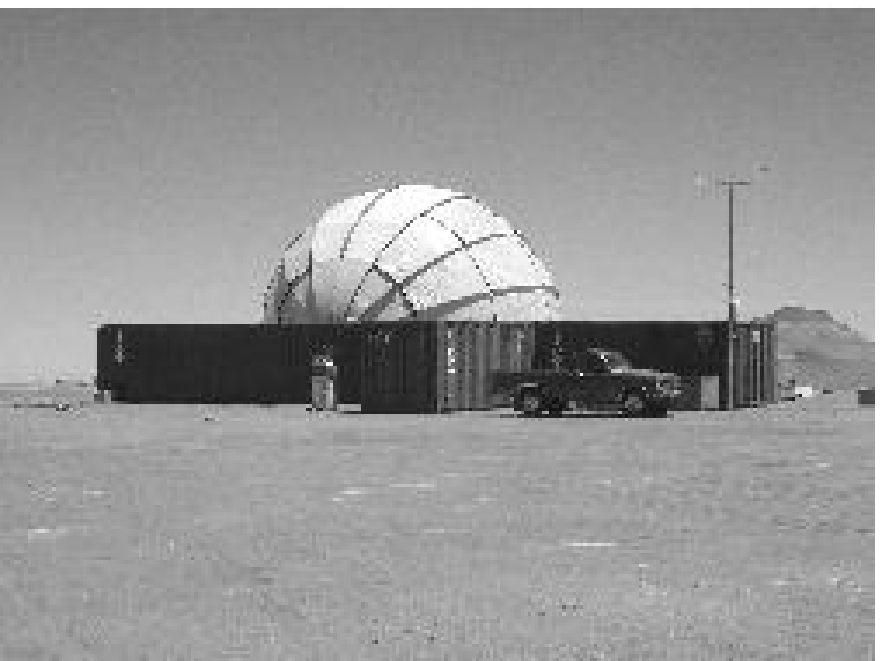}

\footnotesize
Fig. 6.---The CBI site with the dome closed.
\normalsize
\vspace{3mm}

\section{Antennas}

The main constraint on the design of the CBI antennas is low cross-talk.
This requires antennas with very little scattering, suggesting an unblocked
design. Corrugated feedhorns have ideal performance characteristics, but are
somewhat impractical for a $\sim $1 m aperture. A lens would be required to
reduce the horn length and loss in the lens would seriously degrade the
sensitivity of the instrument. Offset reflectors are also an obvious choice,
but they are difficult to close-pack and access to the receivers is awkward.
Because of these problems, we used the shielded cassegrain design shown in
Fig. 7. This has a 0.9-m diameter $f/0.33$ machined cast aluminum primary,
and all 13 primaries made for the CBI have very small surface profile
errors. Based on 6298 measurements, evenly spaced across the surface of each
primary, using a Zeiss Coordinate Measuring Machine, the deviation from a
paraboloid of focal length 0.3002 m ranges from 110 $\mu $m rms ($\lambda /88
$ at 31 GHz) for the first 3 primaries made, down to 30--40 $\mu $m rms ($%
\lambda /322$--$\lambda /242$ at 31 GHz) for the last 10 primaries. The 0.155-m
diameter secondary is made of carbon fiber epoxy and weighs only $\sim $80
g. It is supported on four feedlegs made of expanded polystyrene. The
feedlegs have a U cross-section, hot-wire cut from 2 lb ft$^{-3}$ expanded
polystyrene stock. During assembly of the antenna, the secondary is
supported on a fixture attached to the primary, and the secondary and
feedlegs are glued in place. The polystyrene feedlegs cause very little
scattering and contribute only $\sim $0.5 K to the system noise. The
cassegrain antenna sits in the bottom of a deep cylindrical shield which
reduces cross-talk due to scattering from the secondary and the cryostat.
Scattering from the rim of the shield is reduced by rolling the rim with a
radius of $\sim $5$\lambda $ (Mather 1981). The shield is made from a sheet
of 1/16'' aluminum, welded into a cylinder and then spun to form the rolled
rim. The height of the shield was chosen so that the rim intercepts the beam
where the electric field is about one tenth of the on-axis field. This
reduces the forward gain of the antenna by $\sim $1\%. Ohmic losses in the
shield, measured using room temperature and liquid nitrogen loads in front
of antennas with shields of different heights, contribute $\sim $0.5 K to
the system noise. A 0.36-mm thick woven teflon window (W.L. Gore Inc.),
attached to the front of the shield, protects the antenna components from
the weather. This window contributes $\sim $0.5 K to the system noise, so
the total antenna contribution is $\sim $1.5 K. The antenna is fed by a
wideband corrugated horn at the cassegrain focus, illuminating the secondary
with a $-11$ dB edge taper. The horn has a semi-flare angle of 15$^{\circ }$
and an aperture of 8.4$\lambda $ at the band center. The $-3$ dB beamwidth of
the horn varies by only $\sim $1\% over the 26-36 GHz band, so the
efficiency of the antenna is essentially independent of frequency. A 15$%
^{\circ }$ semi-flare angle minimizes the horn aperture for approximately
constant beamwidth over the band (Clarricoats and Olver 1984). This in turn
minimizes the diameter of the secondary, but the blockage is still quite
high at $\sim $2\% and this degrades the efficiency of the antenna by $\sim $%
4\% (Kildal 1983).

The measured half-power beamwidth of a CBI antenna varies from 51' at 26 GHz
to 38' at 36 GHz and the first sidelobe is $\sim -18$ dB. Offsets between
the antenna boresights are a few arcminutes and are determined primarily by
machining tolerances in the feedhorn mounts. The antenna beams can be
aligned by adjusting the receiver mounts, but usually we just measure the
beam for each baseline and band. The cross-talk between a pair of touching
antennas has a peak value of $-100$ dB at the low frequency end of the band,
decreasing to $-115$ dB at the high frequency end (Padin et al. 2000). This is 
$\sim $30 dB better than for antennas without shields. In any 1 GHz CBI
band, the average cross-talk is at most $-110$ dB, so the maximum false signal
is 6 $\mu $K with 20 K receivers. The actual false signal will be smaller,
because the noise emitted from a receiver input is not completely correlated
with the receiver noise. Unfortunately, a direct measurement of the false signal
is difficult, because ground pickup dominates the correlator output for the short
baselines in the CBI. Between the feedhorn and receiver, there
is a rotating half-wave plate phase shifter which is included to provide
additional rejection of the cross-talk between antennas (and is also useful
for identifying unwanted signals which do not enter through the feedhorn).
If all the phase shifters in the array are rotated synchronously, the
correlator output due to signals from the sky does not change, but the phase
of the contribution due to cross-talk changes by twice the phase change
introduced by each phase shifter. At the band center, the phase shifters
have a range of $\pm 2\pi $ and rotating all the phase shifters through $N$
turns in an integration period completely rejects the cross-talk. At the
band edges, a single turn per integration reduces the correlator output due
to cross-talk by about a factor 10. Any stable residual false signals can be
measured by rotating the entire array about the optical axis.

\vspace{3mm}
\epsfxsize=220pt
\hspace{4pt}
\epsffile{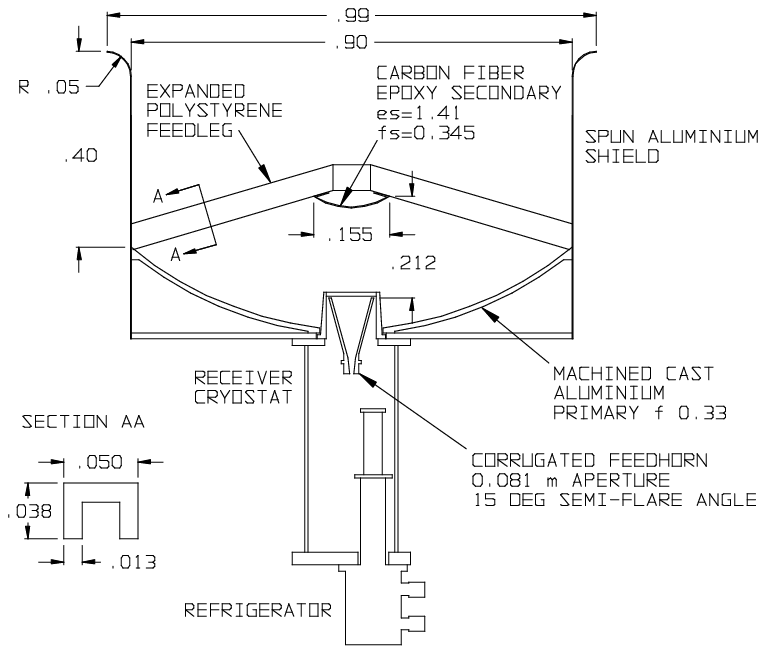}

\footnotesize
Fig. 7.---Cross-section of a CBI antenna. Dimensions are in meters.
es and fs are the eccentricity and focal length of the secondary. The top of
the cryostat is bolted to the back of the antenna and three tensioned struts
run from the bottom of the cryostat to the antanna platform to stiffen the
assembly.
\normalsize
\vspace{3mm}

\begin{figure*}
\figurenum{8}
\epsfxsize=524pt
\epsfbox{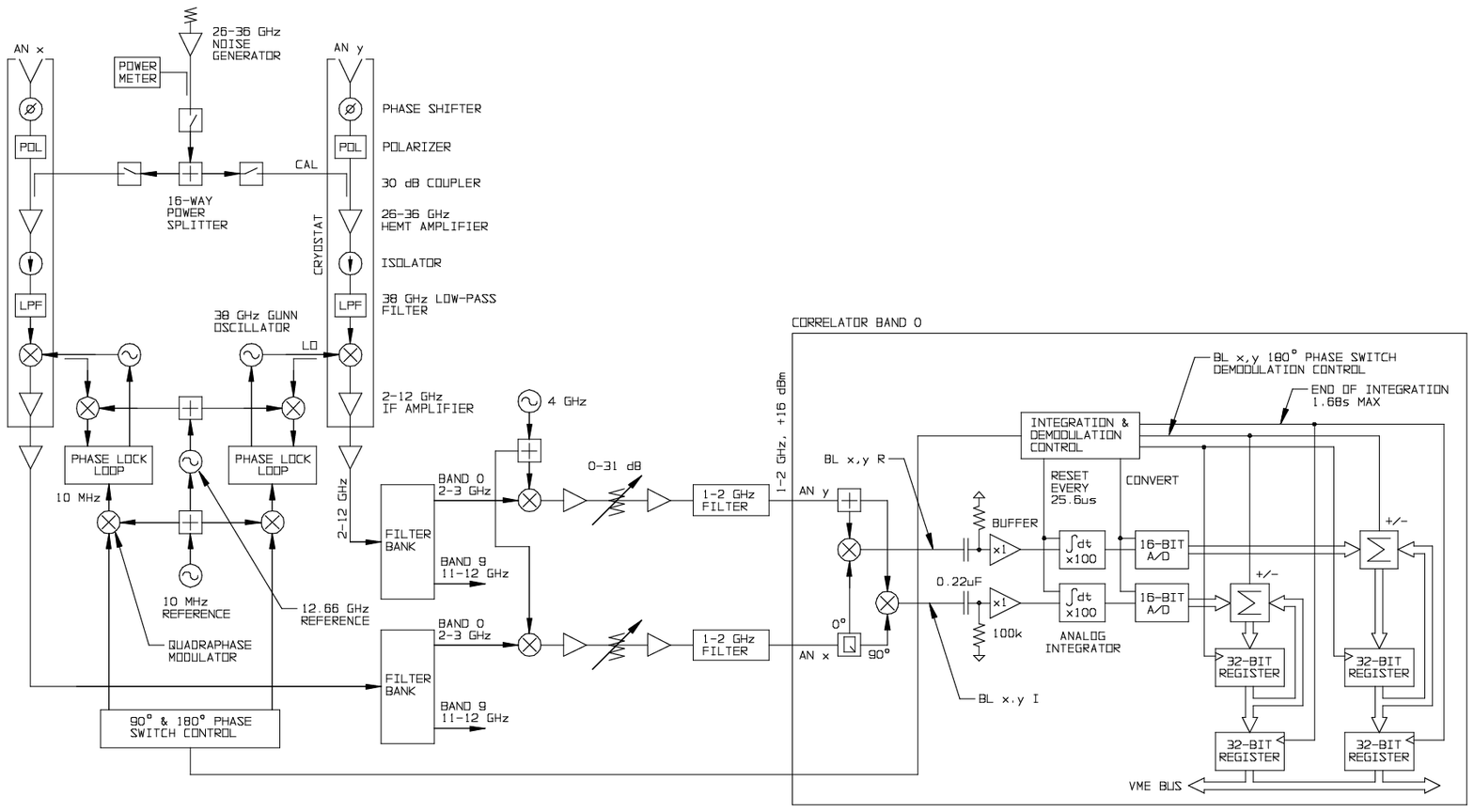}
\caption{Signal processing in the CBI. Receiver details are shown
for 2 antennas, x and y, and for band 0 of the downconverter. Correlator
readout details are given for just one baseline (x, y) in band 0. + is a
power splitter and Q is a quadrature hybrid.}
\end{figure*}

Most of the possible CBI array configurations have many $\sim $1 m
baselines. These are most sensitive to emission on $\sim $30' angular
scales, so the Sun and Moon cause serious problems for the CBI. The Sun is
so bright that it completely dominates the CMBR signal, and the CBI is
restricted to night-time observations. The Moon can be a problem if it is $<$%
60$^{\circ }$ from the optical axis of the telescope. At 60$^{\circ }$, the
primary beam response is $\sim -70$ dB and the Moon gives a $\sim $10 $\mu $%
K signal on a 1 m baseline. Emission from the ground is also a problem and
is particularly serious if the fringe pattern for a short baseline stays
roughly parallel to the horizon as the array tracks the target field. In
this case, the ground signal at the correlator output can 
be $\sim $1 mK and may be roughly constant
for tens of minutes. The ground signal is very repeatable, even on
timescales of several hours, so we can measure the difference between two
adjacent fields to remove the ground contribution. Typically, we observe the
target field for 8 min and then switch to a trailing field at the same
declination, but 8 min later in right ascension. The two fields are observed
over exactly the same azimuth-elevation track, so they have identical ground
contributions. This approach gives difference maps and reduces the
sensitivity of the instrument by $\sqrt{2}$ for CMBR\ observations and 2 for
other observations. (The sensitivity degradation is less for CMBR
observations because both the target field and trail field have CMBR
fluctuations and the rms of the difference between the fields is $\sqrt{2}$
larger than the rms for either field.) For observations of the
Sunyaev-Zel'dovich effect, CMBR fluctuations in the trail field are a
serious contaminant, so we also observe a lead field and take the difference
between the target field and the average of the lead and trail fields. This
reduces the sensitivity by a factor $3/\sqrt{2}$. Ground signals also
degrade the system noise temperature. At the zenith, the spillover
contribution is only $\sim $1 K, but this increases to $\sim $5 K at 40$%
^{\circ }$ elevation, because ground emission reflects off the inside of the
antenna shield cans and enters the receiver feedhorns. We could improve the
sensitivity of the CBI by building a ground shield, but the required
structure is $\sim $10 m high and $\sim $25 m in diameter. The cost of a
rigid structure of this size is prohibitive and the high winds and heavy
snowfalls we experienced during our first year of operation make a flexible
shield impractical.

\section{Signal processing}

Each antenna in the CBI has a cooled heterodyne receiver with a 26--36 GHz
InP HEMT amplifier, Schottky mixer and a 2--12 GHz IF amplifier. The receiver
details are shown in Fig. 8. All the receiver components, including the
feedhorn, are cooled to $\sim $8 K using a 2-stage Gifford-McMahon
refrigerator (APD Cryogenics Inc.). The 1st stage of the refrigerator is at $%
\sim $40 K and cools the cryostat radiation shields and heatsinks for the
receiver wiring. Compressed helium for the refrigerators is supplied by 7
scroll compressors located on the telescope azimuth platform, with each
compressor driving 2 refrigerators. The 26 helium lines (for 13 receivers)
run through both the elevation and parallactic angle cable wraps.

Between the rotating half-wave plate phase shifter at the feedhorn output
and the HEMT amplifier input, there is a quarter-wave plate circular
polarizer with a mode suppressor. The half and quarter-wave plates are
tapered teflon slabs glued into 8-mm diameter circular waveguide (Ayers
1957). Deviations from quarter-wave are $\sim $5$^{\circ }$ at the band
edges, which corresponds to a polarization leakage factor of $\sim $0.1. The
mode suppressor is a tapered piece of mica resistance card, glued into small
broached slots in the circular waveguide, with the card parallel to the
unwanted electric field. This attenuates the unwanted polarization by $\sim $%
40 dB, reducing reflections between the input of the HEMT amplifier and the
antenna which increase the response of the receiver to the unwanted
polarization. The rotating half-wave plate runs in ball bearings and has
chokes at the interface between the rotating and fixed sections of
waveguide. The rotating section is positioned by a stepper motor and
encoder, which are inside the cryostat but at room temperature. The
insertion loss of the complete half and quarter-wave plate and mode
suppressor assembly is $\sim $0.15 dB at 8 K.

Cooling the feedhorn imposes tight constraints on the design of the cryostat
window because the feedhorn mouth is in front of the primary and the size of
the cryostat surrounding the feedhorn sets the minimum blockage for the
antenna. The cryostat window must be thin, so that the cryostat wall is as
far away as possible from the secondary, but the window must be strong
enough to support atmospheric pressure and must attenuate infra-red
radiation enough to prevent excessive heat loading. These constraints are
difficult because the window is $\sim $10 cm in diameter. We use a 0.05-mm
thick mylar film stretched and glued between two thin stainless-steel rings
(like a drum skin) to support a 2lb ft$^{-3}$ expanded polystyrene infra-red
block. The infra-red block is 1 cm thick at the edge and the inside face is
tapered to increase the thickness to 2 cm at the center, to prevent the
polystyrene from cracking under load. The vacuum window is 3-mm thick
closed-cell expanded polyethylene foam. This rests on the flat outside face
of the infra-red block and is clamped directly to the cryostat wall without
an o-ring seal. The wall of the cryostat and the vacuum window retaining
ring are very close to the feedhorn mouth, and to reduce scattering, the
inside surfaces are grooved to form a corrugated circular waveguide
extending $\sim $1.5 cm in front of the feedhorn.

The sensitivity of the CBI is set primarily by the noise temperature of the
26--36 GHz HEMT amplifiers. These are 4-stage amplifiers with 100 $\mu $m InP
HEMTs, giving $\sim $30 dB gain and amplifier noise temperatures in the
range 13 K, at the center of the band for the best amplifier, to 25 K, at
the edge of the band for the worst amplifier (Pospieszalski et al. 1994,
1995). The cooled downconverter contribution to the receiver noise varies
from $\sim $2 K at 26 GHz to $\sim $0.5 K at 36 GHz. Loss in the
phase-shifter and polarizer add $\sim $1 K, the antenna contribution is $%
\sim $1.5 K and ground spillover at the zenith $\sim $1 K. Under good
observing conditions, the atmosphere at the Chajnantor site contributes $%
\sim $1 K, so with 2.7 K from the CMBR, the system noise temperature is in
the range 21--34 K.

Signals from the 13 antennas in the CBI are cross-correlated in an analog
correlator with a filterbank architecture (Padin et al. 2001b). The analog
filterbank approach was chosen to keep the cost and size of the system down
and to give high correlator efficiency, but it does require a calibration
scheme to measure gain and phase errors in the multipliers. The 2--12 GHz
signals from the cooled downconverter in each receiver are split into ten
1 GHz bands by a bank of filters. Each 1 GHz band is then downconverted to
1--2 GHz, which is the operating band for the multipliers. The details of the
downconversion scheme are shown in Fig. 8.

Each baseline in the array requires a complex multiplier to measure the
cross-correlation for each band. This involves distributing many 1--2 GHz
signals between antennas and multipliers, and finding an efficient
architecture was one of the key design challenges in the CBI. The CBI
correlator uses the square array of multipliers shown in Fig. 9 to measure
all the cross-correlations for a band. In this scheme, the top right half of
the array measures the real parts of the cross-correlations, and the bottom
left half measures the imaginary parts. The array has just one quadrature
hybrid per antenna, so the circuit is very compact. Signals from the
antennas are distributed by a grid of microstrip transmission lines, with a
4-GHz bandwidth Gilbert Cell multiplier (Gilbert 1974) at each crossing
point. The signals are sampled by 800 $\Omega $ tap resistors, which form
potential divider circuits with the 50 $\Omega $ input impedance of the
multiplier chips. The multipliers are linear for input powers below 
$\sim -15$ dBm, but at $\sim -20$ dBm, noise from the post-multiplier amplifiers
degrades the efficiency of the correlator by $\sim $2\%. Thus, we can
tolerate a power variation of only $\sim $3 dB over the multiplier array,
which requires tap resistors $\geq $800 $\Omega $. A larger tap resistance would
give smaller power variations over the array, but at the expense of higher
input power. With 800 $\Omega $ tap resistors, the input power is +16 dBm,
which is easy to provide. The multiplier array is made entirely of chip
components, glued to a 97 x 69 mm substrate, with 25-$\mu $m diameter
wirebond connections. The grid spacing is 5 mm; set primarily by the size of
the coupling capacitors at the multiplier inputs and the pins which take the
multiplier output signals through the bottom of the multiplier box. This
puts the first Bragg reflection at $\sim $10 GHz, which is well above the
operating band of the multipliers. The 5 mm grid spacing leaves very little
space for the quadrature hybrids and power splitters in the signal
distribution, so these are realized using lumped elements. The power
splitters have a 3-pole low-pass filter in each arm and the quadrature
hybrids comprise a power splitter followed by $+45^{\circ }$ and $-45^{\circ
}$ phase-shift networks. Sensitivity degradation due to passband errors in
the correlator signal distribution is just a few percent. (A 3 dB p-p
passband error causes a 5\% degradation in sensitivity (Thompson and
D'Addario 1982) and typical passband errors in the power splitters and
hybrids are $\sim $0.2 dB p-p and $\sim $1 dB p-p, respectively.) The array
of multipliers also includes tunnel-diode total power detectors for each
antenna. These are used to set the correlator input power, to ensure the
correct operating point for the multipliers, and for receiver noise
temperature measurements using hot and cold loads.

\vspace{3mm}
\epsfxsize=252pt
\hspace{-13pt}
\epsffile{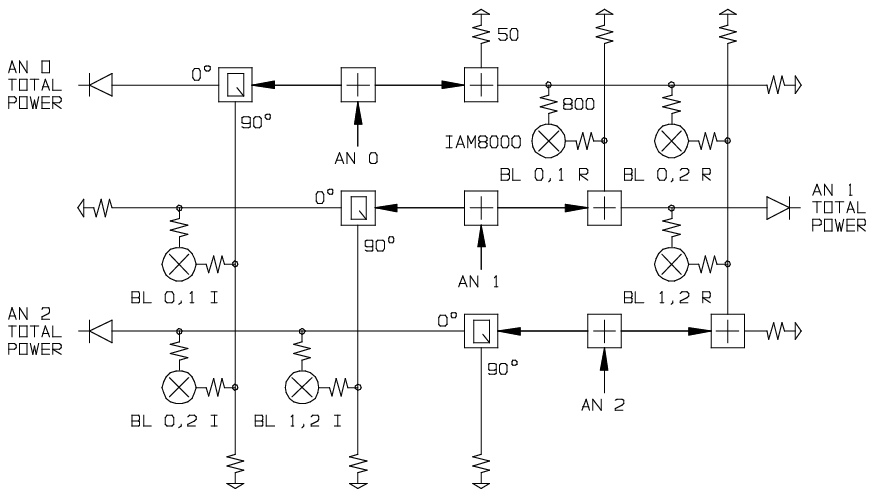}

\footnotesize
Fig. 9.---Architecture of the CBI correlator. This diagram shows a
multiplier array for just 3 antennas. The CBI correlator uses a similar, but
larger, array with 13 antenna inputs and 156 multipliers. The antenna inputs
are along the diagonal, and for a square array, the path lengths to the 2
inputs of a multiplier are the same.
\normalsize
\vspace{3mm}

For observations of CMBR anisotropies, we cannot tolerate false signals
bigger than a few $\mu $K, so with $\sim $25 K system noise temperatures,
the effective cross-talk between the multiplier inputs in the correlator
must be $<-140$ dB. This limit applies to all sources of cross-talk e.g. in
the receiver and downconverter local oscillators, IF amplifiers and in the
correlator itself. In most parts of the instrument, low cross-talk can be
achieved easily, but the performance of the correlator is limited by
cross-talk in and between the multipliers, which can be as much as $\sim -40$
dB. False signals due to cross-talk in the signal processing electronics,
and mains pick-up at the multiplier outputs, are rejected using a 180$%
^{\circ }$ phase-switching scheme in which the receiver local oscillators
are inverted in Walsh function cycles (Urry et al.1985). This modulates the
sign of the multiplier outputs for the desired signals and demodulation
rejects the false signals. In the CBI, the Walsh functions have a maximum
sequency of 32 and a period of 25.6 $\mu $s (i.e. 0.82 ms for a complete
32-state cycle), so even quite rapidly varying false signals are strongly
rejected. Each multiplier output has a high-pass RC filter which removes the
multiplier DC offset and any DC components generated by cross-talk. This is
followed by a 25.6 $\mu $s analog box-car integrator and a 16-bit
analog-to-digital converter. The subsequent signal processing, which
includes the Walsh function phase-switch demodulation and integration for up
to 1.68 s, is entirely digital and is handled by an array of
field-programmable gate arrays. The CBI correlator, including power supplies
and water to air heat exchangers for cooling, occupies about half of a 12U
VME crate.

The efficiency of the CBI correlator, measured using 1--2 GHz noise sources
to simulate the receivers and the signal, is in the range 0.9--1. The
efficiency of the entire receiving system, measured with room temperature
loads at the receiver inputs and the noise calibration source in Fig. 8 as a
correlated signal, is in the range 0.8--1, depending on the passband errors
for the channel. A typical image made with the instrument has a noise level
within $\sim $20\% of the thermal noise. False signals due to cross-talk in
the instrument are at the level of just a few $\mu $K. In a 3-receiver test,
with room-temperature loads at the receiver inputs, the average of the 3
real and imaginary channels was 0.6x10$^{-7}$, in units of correlation
coefficient, after integrating for 20 hrs. The largest correlator output, in
a real or imaginary channel, was 2.2x10$^{-7}$. This is $\sim $3 times the
rms, so it is still consistent with a mean correlator output of zero, but if
it were a residual false signal, the level would be just 6 $\mu $K with 25 K
system noise temperatures. The residual false signal can be removed by
rotating the antenna platform during an observation, or by measuring the
difference between 2 fields (see Sec. 8). Quadrature errors in the
correlator are typically $\sim $1 dB and $\sim $5$^{\circ }$, with a
worst-case of 3 dB and 15$^{\circ }$. The quadrature errors must be measured
in order to calculate the actual cross-correlation from the correlator
outputs (Padin et al. 2001b). This is done by injecting a correlated signal
from the noise calibration source in Fig. 8 and sequentially changing the
phase of each receiver local oscillator by 90$^{\circ }$. The receivers have
phase-locked Gunn local oscillators with a quadraphase modulator in the
reference for the phase-lock loop. The modulators are used for correlator
quadrature error measurements and for the 180$^{\circ }$ Walsh function
phase-switching scheme. A quadrature error measurement takes $\sim $4 min
and we typically do this twice per day. Variations in the quadrature errors
are $\sim $2\% and 1$^{\circ }$ p-p on timescales of a day to a month.

All the signal processing electronics are mounted on the antenna platform to
eliminate instability due to moving cables. The electronics boxes are also
temperature controlled to $\sim \pm $1 K using water to air heat exchangers.
Gain and phase variations in the instrument are measured every $\sim $10 min
using the noise calibration source, which is in turn calibrated using daily
observations of planets, quasars, radio galaxies and supernova remnants. The
calibration noise is injected as early as possible in the receiving system
so that only the feedhorns, phase-shifters and polarizers are outside the
measurement loop. A diode power sensor measures variations in the output of
the noise calibration source and the noise distribution components have low
temperature coefficients and are temperature controlled to $\sim \pm $3 K.
(The noise distribution and local oscillator reference cables run in plastic
conduits along with a cooling water hose.) Variations in the calibration of
the CBI, due to variations in the noise distribution components, are at the
1\% and 1$^{\circ }$ level. Since the injection point for the calibration
noise is ahead of the receiver local oscillators, phase-switching offers no
rejection of cross-talk between the receivers through the noise calibration
source. The diode switches at the outputs of the 16-way power splitter in
Fig. 8 are included to give high isolation between the receivers when the
noise is off, but these switches limit the stability of the calibration
scheme.

\section{Control and monitoring}

\vspace{3mm}
\epsfxsize=252pt
\hspace{-13pt}
\epsffile{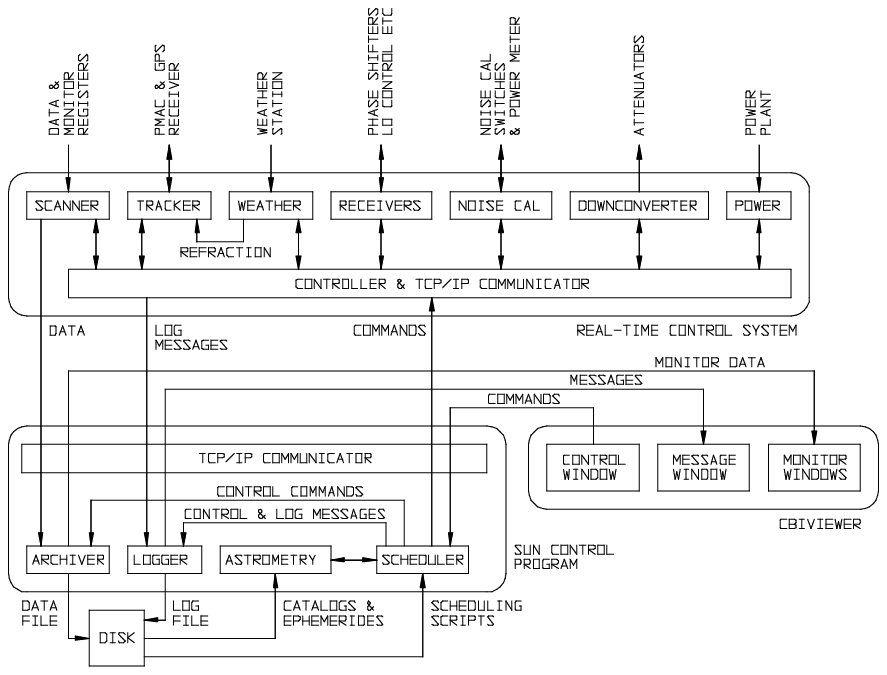}

\footnotesize
Fig. 10.---The CBI control system.
\normalsize
\vspace{3mm}

Fig. 10 shows a block diagram of the CBI control system. Telescope control,
data collection and monitoring functions are handled by a real-time control
system running under VxWorks (Wind River Systems Inc.) on a Motorola 68060
computer. The scanner task reads all 1690 correlator channels (156
multipliers plus 13 total powers for 10 bands) and about the same number of
monitor points (e.g. receiver temperatures and telescope position) typically
every 0.84 s. The tracker calculates the required telescope position every
second and the other tasks control the instrument set-up (e.g. the positions
of the phase shifters in the receivers and the settings of the attenuators
in the downconverter). A control program running on a dedicated Sun
workstation schedules operation of the real-time control system, manages
source catalogs and ephemerides, archives data and handles communications
with ``cbiviewer'' user interface programs. Cbiviewer includes a terminal
window for entering control commands and submitting observing schedules and
provides a powerful graphical interface for monitoring data and the state of
the CBI. Multiple cbiviewer connections are allowed so the instrument can be
controlled and monitored from different locations simultaneously. For most
observations, we integrate the data for 8.4 s before archiving, but for fast
operations, such as noise calibrations and optical pointing measurements, we
archive 0.84 s integrations. The data rate is typically 100 MB/day and we
archive on 4 GB magneto-optical disks. The magneto-optical disks are used to
transport data from the CBI site. We also have a modem connection via a
cellphone, which permits remote monitoring of the weather station, power
plant and cryogenics. The CBI can operate unattended, except for opening and
closing the dome, and cbiviewer has a programmable alarm which summons the
observer if the telescope stops tracking or if a monitor (e.g. wind speed)
is out of range.

\section{Observations}

\vspace{3mm}
\epsfxsize=250pt
\hspace{-12pt}
\epsffile{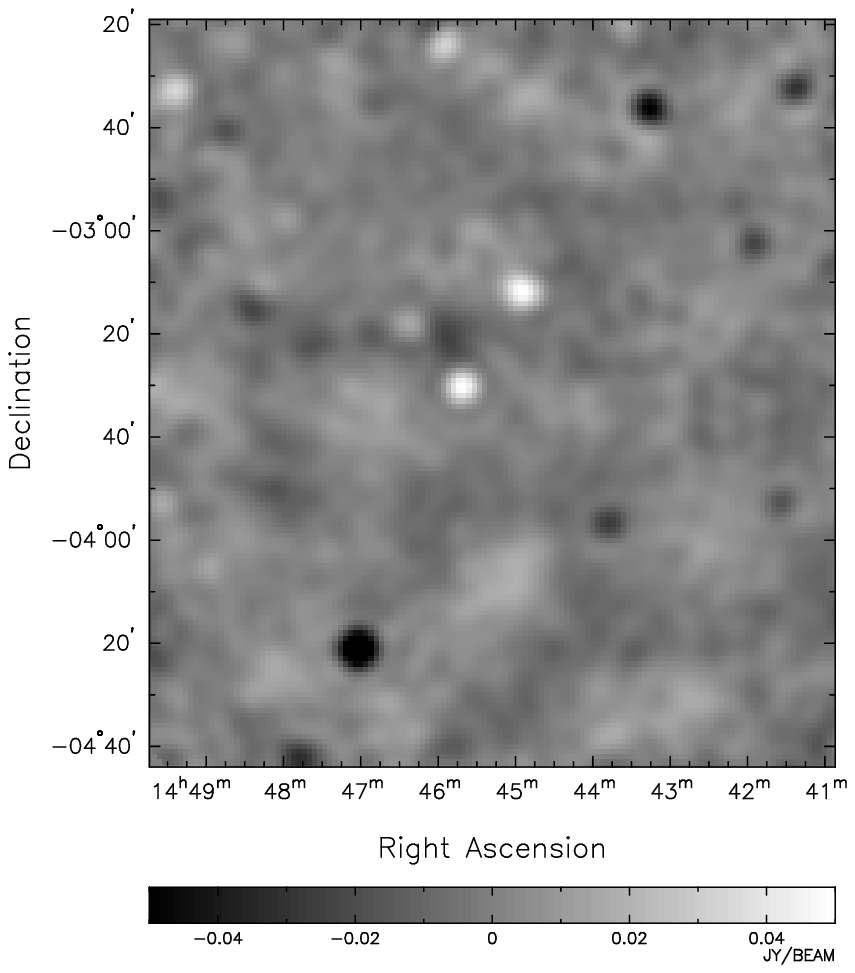}

\footnotesize
Fig. 11.---Mosaic of the 14$^{h}$ CBI field made from 42 main-trail
observations ($\sim $200 hrs total observing time). All 10 frequency bands
have been combined and the individual maps for each pointing center have
been corrected for their primary beams. The mosaic was made pixel-by-pixel
by summing the individual maps, weighting by the inverse variance. The rms
noise in the mosaic varies from 1.5 mJy/beam in the central area to $\sim $5
mJy/beam at the edges. The synthesized beamwidth varies from 4.7' to 5.3'
(FWHM). The brightest point source in this image has a flux density of $-96$
mJy.
\normalsize
\vspace{3mm}

Our current CMBR observations are in three 2$^{\circ }$ x 2$^{\circ }$ fields
separated by $\sim $6 h in right ascension. The fields were chosen to have
IRAS 100 $\mu $m emission $<$1 MJy sr$^{-1}$, low synchrotron emission and
no point sources brighter than a few hundred mJy at 1.4 GHz. Each night, we
observe the two fields that are visible, along with several amplitude and
phase calibrators. Virgo A, Taurus A and Jupiter are our primary amplitude
calibrators. To permit subtraction of the ground contribution (and any
constant false signals generated in the instrument) observations are broken
up into scans with 8 min on the main field, and 8 min on a trailing field at
the same declination, but 8 min later in right ascension. The main and trail
fields are observed over the same azimuth-elevation track and the difference
between the two fields shows no evidence of ground contamination. For
example, dividing the observation into rising and setting halves yields
similar images (with similar noise outside the primary beam) for the
differenced fields, even though the main and trail field images show quite
different ground contributions for the two halves of the observation (Padin
et al. 2001a). The noise calibration source is fired for 10 s at the
beginning and end of each 8 min scan and just before each calibration
observation. Gain and phase variations during a 6 h observation are at most
10\% and 5$^{\circ }$ p-p before the noise calibration is applied, and
typically 1\% and 1$^{\circ }$ p-p after calibration. During an observation,
the array continuously tracks the parallactic angle and is rotated 20$%
^{\circ }$ or 30$^{\circ }$ after each main and trail scan pair. This
improves the synthesized beam and permits polarization measurements. We
build a mosaic on a grid of pointing centers spaced 20' and observe for one
night on each grid point. Fig. 11 shows a mosaic of the 14$^{h}$ CBI field
made from 42 pointings. The extended structures in this image are
fluctuations in the CMBR. They are most obvious on scales of a few tens of
arcminutes because the CBI has low sensitivity to larger angular scales and
the power spectrum of CMBR fluctuations falls off rapidly on smaller angular
scales. Since the image is a difference between main and trail fields, point
sources can appear either positive or negative.

\vspace{3mm}
\epsfxsize=250pt
\hspace{-12pt}
\epsffile{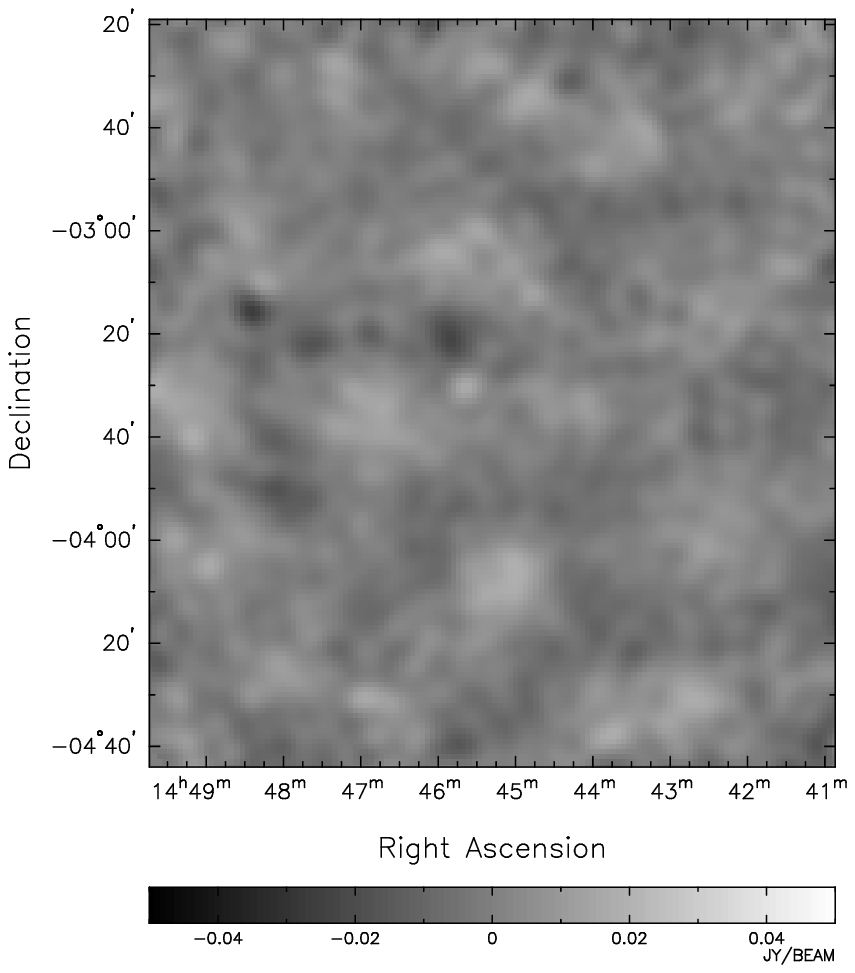}

\footnotesize
Fig. 12.---The 14$^{h}$ field after removal of 63 point sources
monitored by the OVRO 40 m telescope.
\normalsize
\vspace{3mm}

An accurate measurement of the primary beam is required for the power
spectrum extraction, and for making mosaiced images. Errors in the beam
area affect the scaling of the power spectrum, while errors in the beam
shape, and hence errors in the window function, can distort the power
spectrum. We measure the primary beam using observations of Taurus A 
on a grid of pointing centers spaced 7'. With an integration time of 30 s
per pointing, the signal-to-noise ratio on boresight is $\sim $300, so the
error on a measured point on the beam profile is $\sim $0.01. The integration
time can be increased with distance from the beam center to improve the
signal-to-noise ratio in the outer part of the beam.
The measured beamwidth varies from 51'
to 38' (FWHM) across the 26--36 GHz band, and in any 1 GHz band, the
standard deviation of the FWHM beamwidths for the 78 baselines in the
array is $\sim $0.3'. For a particular baseline, the offset of the beam
center is typically a few arcminutes and varies by $\sim $1' across the
26--36 GHz band. Some of these variations with baseline and frequency are
real changes in the beams, so 0.3' and 1' are upper limits to the errors
on the beamwidth and offset.

Flux density measurements of point sources in the CBI fields brighter than $%
\sim $6 mJy at 31 GHz are made using the OVRO 40 m telescope. These
measurements are made within a few days of the CBI observations because many
of the sources are variable. We monitor sources in the NRAO VLA Sky Survey
(Condon et al. 1998) with flux densities $\geq $6 mJy at 1.4 GHz. The very
weak and inverted-spectrum sources that we miss do not seriously contaminate
CMBR power spectrum measurements for $l<1500$. At higher $l$, we must make a
statistical correction for the unmeasured sources based on number counts
from deep VLA observations of the 8$^{h}$ CBI field. For $l<2000$, we can
use the longer baselines in the CBI to monitor point sources. The advantage
of this approach is that the CMBR and point source observations are
simultaneous and have the same pointing and flux scale errors. Fig. 12 shows
the 14$^{h}$ field after removal of the point sources monitored by the OVRO
40 m telescope. Residual point source contamination in this image is small,
indicating that the flux density scales for the CBI and the OVRO 40 m are
commensurate. The good agreement between the CBI and OVRO 40 m point source
measurements is an important demonstration that both systems are working
correctly.

The power and frequency spectra of fluctuations in our fields are extracted
using a fit to the calibrated visibilities, after removal of the bright
point sources. The CBI power spectrum measurements in Fig 1 (Padin et al.
2001a) are from a total of 150 hrs of observations of two main-trail pairs,
one in the 8$^{h}$ CBI field and one in the 14$^{h}$ field. These
observations were made with a ring configuration which did not have enough
baselines in the 1--2 m range to allow us to break the degeneracy between the
angular power spectrum and the frequency spectrum. Subsequent observations
of the 14$^{h}$ field with a fairly compact array give a spectral index of $%
\beta =0.0\pm 0.4$ at $l\sim 600$, so the signals we are measuring are
consistent with blackbody emission and any contamination from foregrounds is
small. The initial results from the CBI are consistent with the $\Omega
_{0}\sim 1$, $\Omega _{b}h^{2}\sim 0.02$, $\Omega _{\Lambda }\sim 0.6$
models implied by DASI (Halverson et al. 2001) and Boomerang (Netterfield et
al. 2001) measurements at lower $l$.

\section{Conclusions}

The CBI has
been making deep single-field observations and 2$^{\circ }$ x 2$^{\circ }$
mosaics for about a year. These observations should yield an accurate
measurement of the power spectrum of CMBR fluctuations in the range $%
500<l<2400$, with an $l$-resolution of $\sim $200 (FWHM). Continuing
deep-field observations and larger mosaics will provide measurements at
higher $l$, and with improved $l$-resolution, allowing us to place strong
constraints on the fundamental cosmological parameters.

During the first year of operation of the CBI we gained substantial
experience working on complex equipment at the Chajnantor site. The biggest
problem in this environment is the wind, and without the protection afforded
by the dome we would not be able to maintain and operate the CBI. Lack of
oxygen is also a serious problem, but portable oxygen systems have been very
effective in enabling us to do strenuous and cognitive tasks. The site is
excellent for CMBR observations when the weather is clear, but snow storms
are frequent and access is often difficult. Pampa la Bola, $\sim $10 km to
the east of the CBI, is more accessible in bad weather and, in retrospect,
would have been a better choice.

In building the CBI we explored unconventional approaches for the mount and
correlator. The CBI demonstrates that good pointing performance can be
achieved with a compact, unbalanced mount, if corrections for the major
structural deformations are applied. We use tiltmeters on the azimuth axis
to measure deformations in the azimuth bearing and linear displacement
transducers to measure deformations in the elevation shaft and encoder
mount. The analog filterbank correlator used in the CBI is a departure from
the current trend towards entirely digital processing in radio telescopes.
The analog system has high efficiency and our wideband noise calibration
scheme ensures that it is stable. It is very compact and fairly inexpensive
and may be appropriate for other instruments which require wide bandwidth
and high sensitivity.

\acknowledgements

This work was supported by the National Science Foundation (award
AST-9802989), California Institute of Technology, Ronald and Maxine Linde
and Cecil and Sally Drinkward.

\end{document}